\documentclass[11pt,twoside]{article}

\begin{document}

\title{A Personal Point of View about Scientific Discussions on Free Will}

\author{S. Esposito\thanks{Salvatore.Esposito@na.infn.it}\\
{\footnotesize{\sl Istituto Nazionale di Fisica Nucleare, Sezione di Napoli, }}\\
{\footnotesize{\sl Complesso Universitario di Monte S.\,Angelo, via Cinthia, I-80126 Naples, Italy}}
}
\date{}
\maketitle

\begin{abstract}
\noindent I briefly present a fairly ``dogmatic'' view about alleged scientific results on free will.
\end{abstract}

\ 

\

\noindent In recent years, an increasing number of papers or discussions about free will within a physics (or, more generally, science) framework have appeared. In particular, a very recent preprint \cite{sabine} summarizes different opinions ``from the perspective of a particle physicist'' and offers further insight about this issue. A ``physical'' model inspired by Quantum Mechanics, as well as a mathematical model based on the von Neumann paradigm, have been formulated \cite{svet}. Moreover, a ``free will theorem'' exists since 2006 \cite{conway}, while a ``free will function'' $F(t)$ has now made its appearance \cite{sabine}, just to quote only few examples of {\it scientification} of the concept. According to Ref. \cite{sabine}, for example, 
\begin{quote}
it is possible to make scientific sense of free will and we have suggested an operational meaning for ``making a choice.'' An agent can make a choice that does not follow from any information available in the past, by reading out the value of a ``free will function'' that has to fulfill the only requirement of not being forward deterministic.
\end{quote}
In order to give an example of what is meant with {\it scientification} of the free will concept, let me quote again from Ref. \cite{sabine}, where the ``free will function'' is introduced:
\begin{quote}
There is a time evolution $H(t)$ that is not forward deterministic in the sense that given the agentÕs state at some time $t_0$, $H(t)$ allows for a set of states at time $t_1 > t_0$. For simplicity, let us assume that the evolution is reversible and deterministic except for a series of moments, $t_i$, $i \in N$, in which the agent ``makes a decision'' and the set of possible states branches into different options that are only probabilistically known. Let us also assume that each decision comes down to choosing one from ten alternatives described by the digits 0 to 9. What we need in order for this evolution to not be random is a function $F(t_i)$ that we can call the ``free will function'' that at any time $t_i$ returns a specific choice, i.e. a digit, and by that selects a uniquely specified path.
\end{quote}
An example of such required function is as well provided:
\begin{quote}
Consider an algorithm that computes some transcendental number, $\tau$, unknown to you. Denote with $t_n$ the $n$-th digit of the number after the decimal point. This creates an infinitely long string of digits. Let $t_N$ be a time very far to the future, and let $F$ be the function that returns $t_{N-i}$ for the choice the agent
makes at time $t_i$. This has the following consequence: The time evolution of the agent's state is now no longer random. It is {\it determined} by $F$, but not (forward) deterministic: No matter how long you record the agent's choices, you will never be able to predict, not even in principle, what the next choice will be.
\end{quote}
I confess to be left completely baffled by the reading of these papers (or similar others), since I have not been able to find in them a proper scientific framework, despite the apparent will of the authors. In my opinion, the point is the following: is it possible to give a scientific framework to the problem about the gender of angels? The answer is certainly: yes. It suffices to define properly (i.e. mathematically) what is an ``angel'' and what is meant for the function termed as ``gender of an angel'', so that the problem is well-defined and, if you are a skilled scholar, then you will be able to solve it. Moreover, if you let the problem to have some relationship with the physical world (to some, desired extent), then you can also be able to embed it into a physical model, possibly with the help of some theorem. This, however, does not mean to make scientific sense to the problem about the gender of angels.

Indeed, I do not question at all about the logical and mathematical correctness of the ``free will theorem'', or the results about the ``free will function'', as well as other similar findings. Simply, I find that such results (and the related discussions) are {\it not scientific}, according to the common meaning that science scholars give to this word. Since Galilei we know that science has to do with experimental observations, and with theories built upon those observations, and later confirmed (or disproved) by further experimental observations. This does not apply at all to discussions about free will. By invoking ``a lot of recent research in neuroscience'' \cite{sabine}, in fact, does not help in providing a genuine scientific basis to the question, since no direct observational results exist yet about the involvement of human brain in this affair (as recognized also in the same paper \cite{sabine}).

More specifically, I do not question about the real existence of free will: simply, we do not have (sufficient) physical evidence for a scientific discussion of it. In the same way, we cannot speak scientifically about angels given the lack of experiments about them, irrespective of their actual existence or not.

Galilei's scientific method can not apply to the present case and, in this respect, the results reported in the literature, even if they were real, are not scientific. This should be remembered when we read about them.


\vspace{0.5cm}



\end{document}